\documentclass[fleqn,usenatbib]{mnras}

\usepackage{newtxtext,newtxmath}
\usepackage[T1]{fontenc}

\DeclareRobustCommand{\VAN}[3]{#2}
\let\VANthebibliography\thebibliography
\def\thebibliography{\DeclareRobustCommand{\VAN}[3]{##3}\VANthebibliography}

\usepackage{graphicx}	% Including figure files
\usepackage{amsmath}	% Advanced maths commands
\usepackage{hyperref}
\usepackage{lineno}
\usepackage{newtxtext,newtxmath}
\title[Automated galaxy classification with Modulos AI]{Image feature extraction and galaxy classification: a novel and efficient approach with automated machine learning}

\author[F. Tarsitano, C. Bruderer, K. Schawinski, W. G. Hartley]{
F. Tarsitano,$^{1}$\thanks{E-mail: federica.tarsitano@phys.ethz.ch}
C. Bruderer,$^{2}$
K. Schawinski,$^{2}$ 
W. G. Hartley$^{3}$\\
\\
$^{1}$Institute for Particle Physics and Astrophysics, ETH Zürich, Wolfgang-Pauli-Strasse 27, CH-8093 Zürich, Switzerland\\
$^{2}$Modulos AG, Technoparkstrasse 1, 8005 Zürich\\
$^{3}$Department of Astronomy, University of Geneva, ch. d'\'Ecogia 16, CH-1290 Versoix, Switzerland
}

\date{Accepted XXX. Received YYY; in original form ZZZ}

\pubyear{2015}

\begin{document}
\label{firstpage}
\pagerange{\pageref{firstpage}--\pageref{lastpage}}
\maketitle

% Abstract of the paper
\begin{abstract}
Machine learning methods are extensively used for a broad range of applications, from healthcare to economy and to natural sciences. They are often used in image recognition and are also employed to classify sequences, which represent a simpler and more convenient format to store and organize data. In this work we explore the possibility of applying machine learning methods designed for one-dimensional problems to the task of galaxy image classification. The algorithms used for image classification typically rely on multiple costly steps, such as the Point Spread Function (PSF) deconvolution and the training and application of complex Convolutional Neural Networks (CNN) of thousands or even millions of parameters. In our approach, we extract features from the galaxy images by analysing the elliptical isophotes in their light distribution and collect the information in a sequence. The sequences obtained with this method present definite features allowing a direct distinction between galaxy types, as opposed to smooth Sérsic profiles. Then, we train and classify the sequences with machine learning algorithms, designed through the platform \textit{Modulos} \texttt{AutoML}, and study how they optimize the classification task. As a demonstration of this method, we use the second public release of the Dark Energy Survey (DES DR2). We show that by applying it to this sample we are able to successfully distinguish between early-type and late-type galaxies, for images with signal-to-noise ratio greater then 300. %reaching an $F_1$ score of $0.89$ for images with signal-to-noise ratio bigger then 300. 
This yields an accuracy of $86\%$ for the early-type galaxies and $93\%$ for the late-type galaxies, which is on par with most contemporary automated image classification approaches.
Our novel method allows for galaxy images to be accurately classified and is faster than other approaches. Data dimensionality reduction also implies a significant lowering in computational cost. In the perspective of future data sets obtained with e.g. {\it Euclid} and the Vera Rubin Observatory (VRO), this work represents a path towards using a well-tested and widely used platform from industry in efficiently tackling galaxy classification problems at the peta-byte scale. 
% In a broader context, this is an interesting machine learning experiment which considers a method already well tested and widely used for several applications and introduces it in Astronomy, with the intention of setting initial ground for future progress in galaxy classification.

\end{abstract}

% Select between one and six entries from the list of approved keywords.
% Don't make up new ones.

%%%%%%%%%%%%%%%%%%%%%%%%%%%%%%%%%%%%%%%%%%%%%%%%%%

%%%%%%%%%%%%%%%%% BODY OF PAPER %%%%%%%%%%%%%%%%%%

\section{Introduction}
Galaxy morphology plays an important role in our studies and understanding of galaxy evolution. Structural components such as bulges, diska, spiral arms and bars formed during galaxies' aggregated formation histories \citep{1981A&A....96..164C, 1996A&A...313...45D, 1996AJ....111.2233E}. As such, morphology is related to other properties that depend on formation and assembly history,
such as colour, stellar-mass and recent Star Formation Rate (SFR) \citep{2004AIPC..743..106B, 2007ApJ...660L..47N, 10.1093/mnras/stz1894}. By looking at the relation between mass and SFR \citep{2007ApJS..173..315S, article, 2021MNRAS.500L..42P}, astronomers have been able to distinguish between three different populations. Most star-forming galaxies belong to the \textit{main sequence}, and present morphological features typical of spiral or irregular galaxies. Objects in this population are also called \textit{Late-Type Galaxies} (LTG). We can identify another population with much lower SFR and different shapes, mostly elliptical or bulge-dominated morphologies: we refer to these as \textit{Early-Type Galaxies} (ETG). The transition between ETG and the \textit{main sequence} is smoothed by an intermediate and less heavily populated region, called the \textit{green valley} \citep{Salim_2014, Schawinski_2014}.

Historically, galaxies were classified as early or late type by visual inspection, with a modern example of classification in this way  provided by \cite{Nair_2010}. Recent \textit{citizen science} projects like \textit{Galaxy Zoo} \citep{LintottKevin, Simmons, Lingard} use the same approach, while benefiting from a huge network of volunteers who are asked to classify galaxies. Quantitative methods for classifying structural properties includes modeling galaxy light profiles with 2D analytic functions, and fitting them to galaxy images. The most commonly used model is the \textit{Sérsic profile} \citep{Sersic}, a parametric function with parameters describing structural properties such as size, magnitude, ellipticity, inclination and the rate at which light intensity falls off with radius (\textit{Sérsic index}). The latter quantifies the concentration of light and it is often used to distinguish between ETG and LTG. In fitting galaxy images, the Sérsic model must be convolved with the Point Spread Function (PSF), in order to account for the seeing and any instrumental distortions of the image. This method is proven to be robust when the multi-dimensional fitting delivers non-degenerate solutions. Alternative non-parametric approaches can be used to analyse the light distribution and quantify the galaxy concentration and level of asymmetry and to search for clumpy regions \citep{Conselice2000}. In this case, the PSF is not taken into account. Several large catalogues of galaxy morphologies exist, based on parametric fitting \citep{Simard2011}, a non-parametric approach \citep{10.1093/mnras/stz1894} or both \citep{catdes}.

Going beyond the traditional methods cited above, machine learning algorithms present an attractive way forward in classifying galaxy images. This has been proven to be successful, as in \cite{Manda2010}, \cite{Dieleman} and \cite{10.1093/mnras/stz2816}, where data sets from \textit{Galaxy Zoo} were used combining human and machine intelligence. 
Supervised CNN have been used also on \textsc{CANDELS} images \citep{Grogin_2011, Koekemoer_2011} to provide galaxy visual classifications \citep{Huertas_Company_2015, 2018MNRAS.475..894T}, to find specific structural features such as bars \citep{10.1093/mnras/sty627} and to classify galaxies according to their bulge+disk composition \cite{2020ApJ...895..112G}. In \cite{2020arXiv200911932C} unsupervised machine learning algorithms on \textsc{SDSS} images \citep{2000AJ....120.1579Y} were explored, foreseeing the application of these techniques on data with higher resolution and deeper depth from the Dark Energy Survey (DES) \citep{2016MNRAS.460.1270D} and the Euclid Space Telescope \citep{Amiaux_2012}. 

The recent increase in the use of machine learning methods has been beneficial for astronomy research, and is of particular interest for extracting information on the evolutionary paths of galaxies from their morphologies. Especially with the exponential rise in the amount of data from modern surveys it has become important to understand and apply intelligent algorithms able to classify galaxies with the same accuracy as human experts, if not even outperforming them. In the works cited above, the CNNs classify galaxy images by processing different levels of information in each layer, aiming at a progressive recognition of complex features. In this approach, image recognition works well if the objects have clear edges. However galaxies' outskirts are smooth: even traditional methods used to measure structural properties, namely the 2D parametric fitting and the non-parametric analyses, are often prone to inaccuracies due to the difficulty of separating galaxy wings from the background. These boundary effects can be mitigated by using model constraints, but cannot completely prevent inaccurate estimations of structural parameters. Machine learning techniques are also subject to mis-classifications for the same reasons, especially with low-resolution images. Another factor to account for when adopting intelligent algorithms is the data management and the speed of the analyses. The increasing volume of available images is difficult to manage and the number of operations processed in CNN models is high. Both training and testing large image data sets requires a lot of time and significant computational costs.
These limiting factors led us to search for an alternative method, which performs an isophotal analysis of the galaxy light distribution, stores the information in a more manageable data format and performs classification lowering the total computational costs. In this paper we show that this method, applied to the second public data release, DR2, \citep{abbott2021dark, 2018PASP..130g4501M} of the Dark energy Survey (DES), can lead to competitive results. Recently \cite{vegaferrero2020pushing} applied CNN to an earlier public release (DES DR1), producing a galaxy morphology catalogue which has an accuracy of $87 \%$ and $73 \%$ for ETG and LTG, respectively, up to magnitude $m_r < 21.5$. In our work we analyse galaxies up to magnitude $m_i<20$, reaching comparable accuracy: $86\%$ for ETG and $93\%$ for LTG.

Section \ref{sec:data} contains more details about the data set. Our method and details on how we perform the isophotal analysis of galaxy images, extract features from their light distribution and collect those into sequences, is described in Section \ref{sec:method}. The sequences are then processed through a neural network designed and run in the framework of Modulos AutoML\footnote{\href{https://www.modulos.ai/}{https://www.modulos.ai/}}. More details are given in Section \ref{sec:framework}. We present the results in Section \ref{sec:results} and discuss further developments in Section \ref{sec:discussion}.

\section{Data}
\label{sec:data}
In this work, we use public images from the Dark Energy Survey \texttt{DR2} release \citep{abbott2021dark, 2018PASP..130g4501M, 2015AJ....150..150F}, available through the public \textit{DES Data Management}.\footnote{\href{https://des.ncsa.illinois.edu}{https://des.ncsa.illinois.edu}} In this section, we provide an overview of the survey, describe the structure of the data set and define the selection function for our sample.

\subsection{The Dark Energy Survey}
\label{sec:DES}
The Dark Energy Survey (DES) is a project aiming to map hundreds of millions of galaxies to measure the effects of dark energy on the expansion history of the Universe and the growth of cosmic structure. The collected data are analysed through different methods: gravitational lensing, galaxy clustering and Baryonic Acoustic Oscillations (BAO). DES used the Dark Energy Camera (DECam) to detect more than 300 million galaxies between the years 2013 and 2019 \citep{Flaugher_2015}. Although conceived for cosmological research, the vast data set assembled by DES represents a powerful survey for the fields of galaxy evolution, stellar populations and Solar System Science too \citep{Abbott}. Moreover, in 2017 DECam provided the optical counterpart of the gravitational wave event GW170817 studied in detail in \cite{Palmese}. The camera has a $2.2^{\circ}$ diameter field of view and a pixel scale of $0.263"$ \citep{Flaugher}. It is mounted on the Victor M. Blanco 4-meter Telescope at the Cerro Tololo Inter-American Observatory (CTIO) located in the Chilean Andes.

\subsection{The data set}
\label{sec:data set}
The DES survey area is covered by images in five photometric bands, \textit{g,r,i,z,Y}. The single exposure images have integration time of 90 seconds in the \textit{g,r,i,z} and 45 seconds in the \textit{Y} band. Data are later processed through the DESDM (DES Data Management) pipeline, which first applies calibrations and coadds the images, then detects and catalogues all the objects in those images \citep{Drlica-Wagner, Morganson2018}. In the image co-addition, the pipeline combines overlapping single-epoch images in one filter and remaps them to artificial tiles on the sky as described in \cite{Sevilla:2011ps}, \cite{Desai} and \cite{Mohr}. Object detection is made using a specific software, called \texttt{SExtractor} \citep{Bertin2011}, which extracts structures from the background and distinguishes between point-like (stars) and galaxies. Then, it performs a photometric analysis, where each object is enumerated and assigned to a set of specific properties, collected in a catalogue. For this analysis, properties of the light distribution are measured, namely the object brightness, quantified in \texttt{MAG\_AUTO}, and its size, called \texttt{FLUX\_RADIUS}, which includes half of the galaxy light. We use these measures to identify and optimise the sample analysed in this work. 

\subsection{Sample selection}
\label{sec:sampleselection}
We apply cuts to the \texttt{SExtractor} catalogues (see below) to select a final sample of 6525 galaxies. We choose objects which are neither truncated nor corrupted or blended to other objects by setting $\texttt{FLAGS}=0,2$. Additionally, we choose bright objects by applying a cut in magnitude. In \cite{catdes}, we observe that robust fits are obtained for objects up to a magnitude of 21.5 in the \textit{i}-band. In order to work with optimal isophotal fitting, in this analysis we make a more conservative cut, setting in the same filter $\texttt{MAG\_AUTO} \leq 20$. For the same reason, we also adopt the cut in signal-to-noise $S/N > 300$. In \cite{catdes} we also flagged those galaxies with size smaller than or comparable to the PSF, because in those cases the PSF significantly affects the way the concentration of light is modelled, leading to degeneracies in the estimation of the size and Sérsic index. Therefore our selection function excludes the galaxies with size smaller than $4 \ px$ in the \textit{i}-band. We also check that the selected objects have physically meaningful measurements, avoiding galaxies with negative or null radii. In processing the data (see Section \ref{sec:method}), we will make use of the \textit{Kron radius}, which is the radius within which approximately $90\%$ of the galaxy light is included. According to the definition in \texttt{SExtractor}, we consider as \textit{Kron radius} the product between the \texttt{KRON\_RADIUS} and the semi-major axis of the galaxy \texttt{A\_IMAGE}. Finally, we exclude from our sample the point-like objects by applying a cut to the \texttt{MODEST} parameter \citep{Drlica_Wagner_2018}. The sample selection is summarized in Table \ref{tab:selection_table}. Additional information about the \texttt{MODEST} star/galaxy classifier and the \texttt{Sextractor} catalogues can be found here: \href{https://des.ncsa.illinois.edu/releases/y1a1/gold}{https://des.ncsa.illinois.edu/releases/y1a1/gold}.

\begin{table}
	\centering
	\caption{Summary of the cuts applied to \texttt{SExtractor} catalogues for the sample selection.}
	\label{tab:selection_table}
	\begin{tabular}{lccr} % four columns, alignment for each
		\hline
		SELECTION TYPE & SELECTION CUT\\
		\hline
		Image flags & \texttt{FLAGS} = 0, 2 \\
		Magnitude & \texttt{MAG\_AUTO} < 20 \\
		%Size (I) & \texttt{FLUX\_RADIUS} > 0 \\
		%Size (II) & \texttt{KRON\_RADIUS} > 0 \\
		S-G & \texttt{MODEST} > 0.005  \\ 
		S/N & \texttt{FLUX\_AUTO/FLUXERR\_AUTO > 300} \\
		\hline
	\end{tabular}
\end{table}

\section{Method}
\label{sec:method}
In this section, we describe our process to transform galaxy images into one-dimensional feature vectors for classification using machine learning. As already mentioned in the introductory sections, this method involves few and fast steps, which is an advantage compared to classification methods that involve several labourious manipulations. More precisely, we refer to two main steps:

\begin{enumerate}
	
	\item production of postage stamp images;
	\item extraction of profiles.
	
\end{enumerate}

We describe the steps below, highlighting the main differences with existing methods.

\subsection{Production of stamps}
For each of our selected galaxies (see Section \ref{sec:sampleselection}), we cut square postage-stamp images
from the relevant DES tiles, with dimensions equal to four times the \textit{Kron radius}. This size is chosen to ensure that the image includes the galaxy light distribution entirely and sufficient non-object pixels to be able to determine the background level. For this operation, we use the publicly available CANVAS algorithm \footnote{\url{https://github.com/Federica24/Cosmo}} (Cut ANd VAlidate Stamps), presented and optimized in \cite{catdes}.

%Given our selected galaxies (see Section \ref{sec:sampleselection}), we cut their stamps from the relative DES tiles. For this operation, we use the publicly available CANVAS algorithm (\textit{Cut ANd VAlidate Stamps}), presented in \cite{catdes} and successfully used in their work. It is publicly available at this link: \href{https://github.com/Federica24/Cosmo}{https://github.com/Federica24/Cosmo}. The size of the stamp is proportional to the \textit{Kron radius} of the galaxy, defined in Section \ref{sec:sampleselection}. We set the half-width of the stamps to twice the \textit{Kron radius} to ensure that the image includes the galaxy light distribution entirely and also some surrounding background. 

\subsubsection{Background}
In standard analyses such as parametric fitting, the background needs to occupy at least $60\%$ of the area of the stamp, in order to obtain a correct fit. The fitting algorithm, in fact, needs to distinguish the light signal from the sky, which becomes challenging towards the outskirts and faint wings of a galaxy. Therefore, a clear separation is only possible if the background occupies a larger area of the stamp than the galaxy \citep{galfit}. In works that perform image classification with neural networks, the preparation of the sample usually includes data-augmentation with image simulations to artificially place a galaxy with well-known classification at different redshifts. This process requires the PSF to be deconvolved and the reconstruction of images with an appropriate percentage of background.
In our work, we do not perform parametric fitting and we extract information solely from the area inside the \textit{Kron ellipse} of the galaxy. Hence our approach is robust to minor defects or mis-estimations of the background or the image stamp size, and we need not be concerned with multiplying the size of our input images to ensure sufficient background coverage.

\subsubsection{Neighbouring objects}
A potential source of errors for the galaxy classification is the presence of neighbouring objects. If not taken into account, both standard fitting algorithms and CNNs are prone to imprecise estimations or classification of the galaxy light profile. We consider two cases:

\begin{itemize}
	\item the neighbour falls outside the \textit{Kron ellipse} of the galaxy;
	\item the neighbour is placed inside the \textit{Kron ellipse} of the galaxy (partially or fully).
\end{itemize}

The first scenario is negligible for our analysis, since we only consider the pixels inside of the \textit{Kron ellipse}. However, we cannot ignore the second case. Our aim is to minimize the number of manipulations applied to the images, so we do not apply any algorithm to identify such cases. Moreover, we would need to distinguish cases that are due to chance alignments from more interesting but possibly similar-appearing cases due to, e.g., galaxy-galaxy mergers or star-forming clumps. This latter task is an avenue of future research for our method, but as contaminants do not change the overall trend of the sequences we obtain for elliptical and spiral galaxies, we do not apply corrections for them in the present work.

\subsection{Extraction of profiles}
\label{sec:extraction_profiles}
\begin{figure}
	\includegraphics[width=\columnwidth]{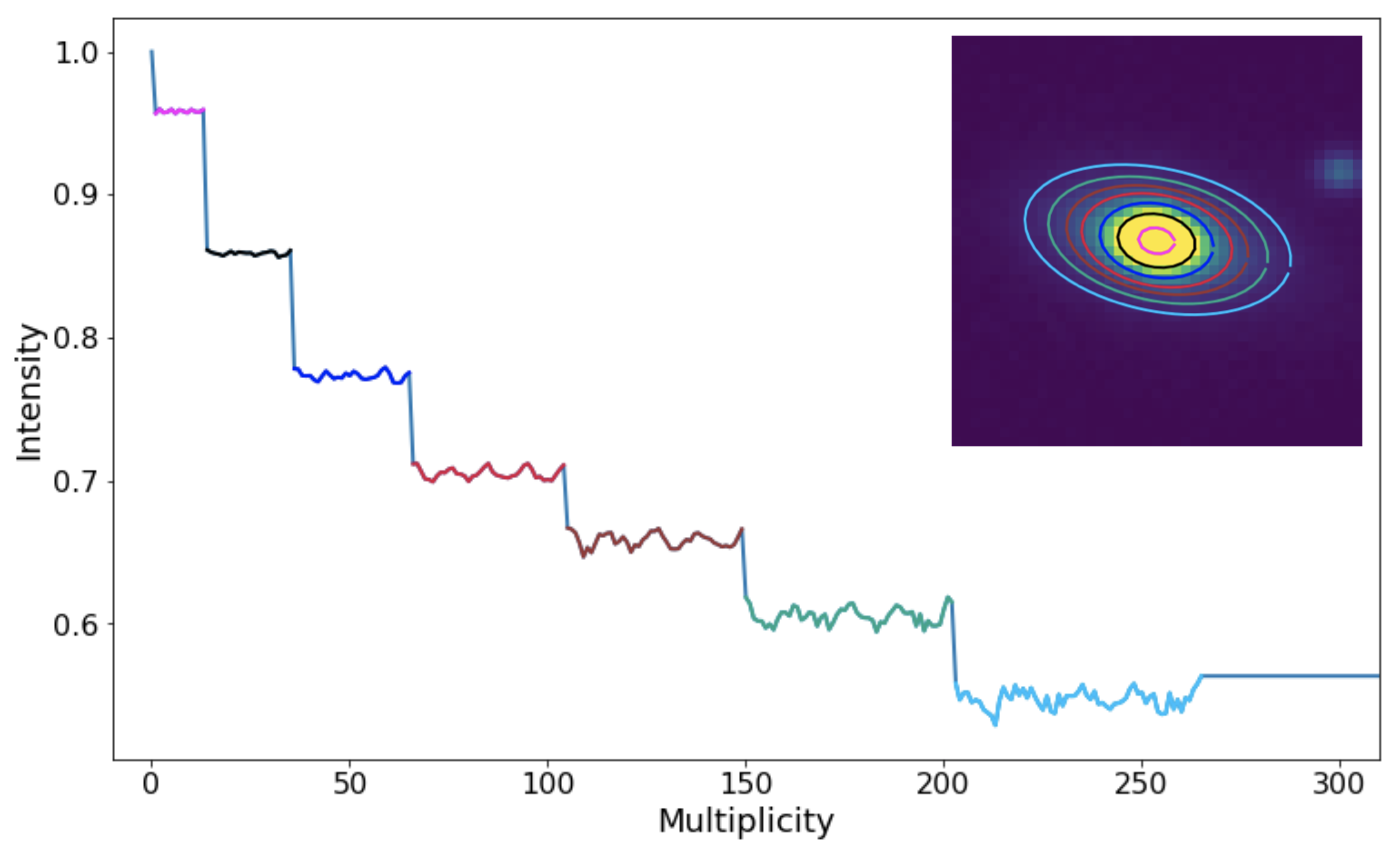}
	\caption{Example of a series extracted from galaxy images and used for classification. The galaxy (an early-type in this case) is processed through an isophote analysis, then the logarithm of the radial intensities of pixels lying on each detected ellipse is read and stored in a series, where the multiplicity axis in the main panel is a counter for the pixel values read. The intensity is from the inner to the outer ellipse. Isophotes and their collected intensities are matched by colour.}
	\label{fig:method}
\end{figure}

The extraction of profiles relies on the elliptical isophote analysis of the galaxy in question. Isophotes are curves connecting locations with the same brightness. We use the algorithm of \textit{Elliptical Isophote Analysis} available in the \texttt{Photutils Astropy} package \citep{photutils}. The algorithm searches for elliptical isophotes iteratively, as described in detail in \cite{isophotes}, up to a user-defined limit, expressed in terms of a maximum value for the semi-major axis of the ellipse. We set this limit to 0.7 times the \textit{Kron radius} in order to exclude the faint wings of a galaxy light distribution from the analysis. Specifically in our fitting routine we observed that such wings result in a noisy tail in the one-dimensional sequences and do not add information useful for distinguishing between different classes of galaxies. Once we measure the isophotes, we proceed by \textit{radially and concentrically} collecting the intensity of pixels falling on the curves, one ellipse at a time. The points collected in this order form our sequence, which we convert to logarithmic scale and normalise such that the brightest pixel has value unity. We can visualize this procedure in Figure \ref{fig:method}, where the isophotes and their radial intensities collected in the series are matched by colour.

\begin{figure*}
	\includegraphics[width=\textwidth]{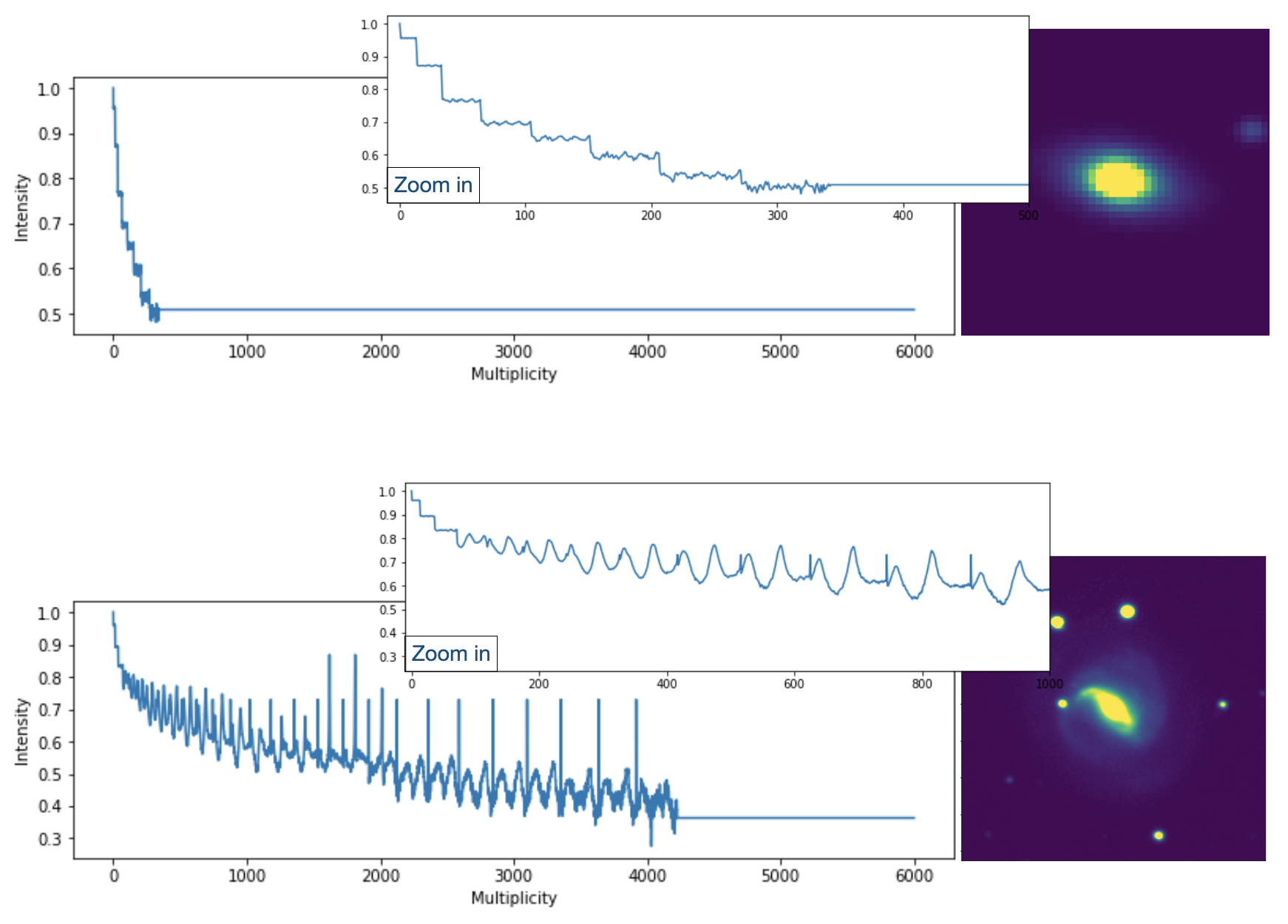}
	\caption{Example of classification between early-type (upper panel) and late-type (lower panel) galaxy, according to their time-series-like profile.}
	\label{fig:example_figure}
\end{figure*}

The series show a clear and expected pattern of decreasing intensity along the x-axis as the isophotes are read from
the centre towards the outskirts of the galaxy. This pattern varies between early and late-type galaxies, as can be seen in \ref{fig:example_figure}. For early-type, mostly elliptical galaxies (upper panel), the trend resembles a step-function, showing regular patterns with decreasing intensity: each step represents an isophote. For late type galaxies, we observe a different trend: in addition to a slower fall off in intensity due to their lower Sérsic index, the presence of spiral arms or clumps adds irregular spikes to the ideal step function. An example is shown in the lower panel of the figure, where we consider a barred-spiral galaxy. 

\section{AI framework and Modulos} 
\label{sec:framework}

We run a \textit{Modulos AI} workflow on a data set randomly split between a test (2175 galaxies) and training+validation sample (4350 galaxies), with all objects visually inspected according to their corresponding 1-D sequences and images. As previously mentioned in Section \ref{sec:method}, we show in Figure \ref{fig:example_figure} examples of sequences for ETG (upper panel) and LTG (lower panel). In this section we describe the properties of the workflow in more detail. We use the Modulos\footnote{\href{https://www.modulos.ai/}{https://www.modulos.ai/}} \texttt{AutoML} platform (version \texttt{0.3.5}) to search for suitable models. The platform is designed to perform automated model selection and training for machine learning tasks, and works in the following way:
\begin{enumerate}
    \item Workflow configuration (ML task): the user selects the data set to be processed, and sets an objective for which it is optimized.
    \item Schema matching: the platform detects the schema of the desired input and output. It then proposes the feature extraction methods and machine learning models applicable to the data set and target objective.
    \item Optimization: using a Bayesian optimizer \citep{srinivas:2009aa}, the platform tries out various combinations of feature extractors, models and their parameters. At each search step, the platform selects a feature engineering method and a model, chooses its architecture and hyperparameters, and trains it. After completing training, the platform uses a validation set to score this particular choice.
    \item End point: There is no clearly defined end point at which the "best model" has been found. However, after a while, the scores for the models begin to converge. As a default, the platform stops if there are no score improvements within 200 steps.
    \item Download: Any trained model can be downloaded and used. We choose the best-performing model.
\end{enumerate}

The key advantages of a platform such as \texttt{AutoML} are that the search for models and configuration is principled and not biased by human intervention. It is also significantly more efficient than optimization searches performed "by hand". During this project, the vast majority of the time was spent on the preparation and the analysis of the data set, with only a few hours required for the automated classification. In our case, we found a suitable solution within 4.30 hours of compute time (14 min 19 sec to train the specific solution). 

The objective we set for optimisation is the $F_1$ score. For multi-class classification, the total $F_1$ score is the unweighted arithmetic mean of the $F_{1,i}$ scores of each class $i$ (macro-averaged). These are the harmonic mean of the precision and recall of the classified samples for each respective class, i.e.
\begin{equation}\label{eq:f1macro}
    F_{1,i} = 2\cdot\frac{\textrm{precision}_i\cdot\textrm{recall}_i}{\textrm{precision}_i+\textrm{recall}_i} = \frac{2\textrm{TP}_i}{2\textrm{TP}_i+\textrm{FP}_i+\textrm{FN}_i},
\end{equation}
where TP$_i$ are the true positives for the classified samples for the class $i$ and FP$_i$ and FN$_i$ are the false positives and false negatives respectively.

\section{Results and discussion}
\label{sec:results}
The automated machine learning framework returns as best solution an XGBoost model with a PCA decomposition as feature engineering method. XGBoost (Extreme Gradient Boosting) is a decision-tree based Machine Learning algorithm using a gradient boosting framework \citep{10.1214/aos/1013203451}. This model reaches a $F_1$ macro score (see eq.~\ref{eq:f1macro}) of $90 \%$ on training data and $89 \%$ on test data, which suggests it is not prone to over or under-fitting. Our best model is publicly available at \href{https://github.com/Federica24/Cosmo}{https://github.com/Federica24/Cosmo} and can be applied to any DES data processed as described in the previous sections.

\subsection{Information extraction}
\begin{figure*}
	\includegraphics[width=\textwidth]{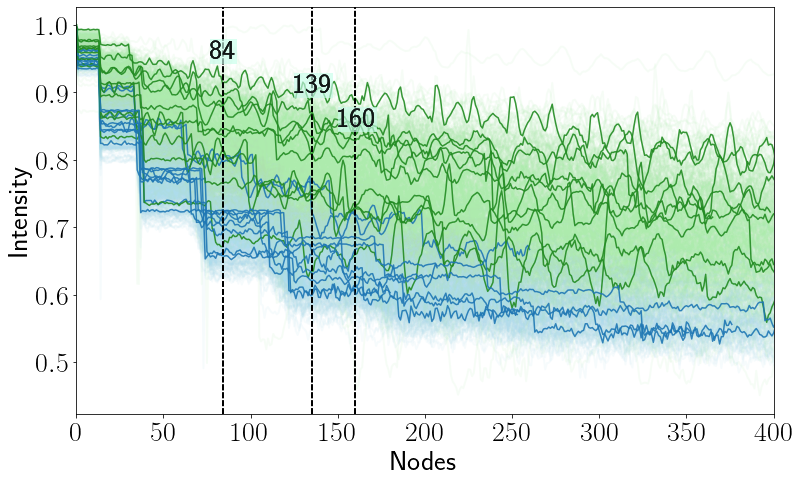}
	\caption{Plot of the sequences corresponding to ETG (in blue) and LTG (in green). Additionally, 10 arbitrary ETG and LTG are drawn. The black, vertical lines indicate the nodes with the largest Gini feature importance.}
	\label{fig:gini}
\end{figure*}

In order to understand why a combination of an XGBoost model and a PCA feature engineering method is found to be performing best, we review the key information accessed during classification by our model and compare it to the structure of our data. In Figure \ref{fig:gini} we show the collection of sequences classified as ETG (in blue) and LTG (in green). Additionally, we have highlighted 10 arbitrary sequences from each class to illustrate the individual profiles. The human eye is able to distinguish between the two classes by looking at the slopes of the sequences (greater for ETG) and comparing the abundance of spike-like features, which correspond to spiral arms of LTG, and to the smoothness of the sequences representing ETG. If spikes occur in the latter, they are more sparse and might refer to the presence of a neighbour (see Section \ref{sec:method} for reference). The information contained in the slopes and features is enhanced by the PCA analysis, which encodes it into a set of components. Each component brings a unique contribution to the automatic classification, giving the features different weight. By computing the Gini feature importance of the collection of decision trees, we can then understand which features contribute the most in predicting a class for a new data point in our model. We show the three most important nodes picked up by the Gini feature importance with black, vertical lines on Figure \ref{fig:gini}. We notice that the three nodes refer to points where the two classes of sequences on average start falling at different rates. In other words, these nodes are the ones with the largest discriminatory power for the collection of decision trees, which perform classifications by sequentially dividing up the sequences. We find that the most important node for our best model (numbered 139) is 15 times more important than the others. This may be because it provides the highest signal-to-noise estimate of the light intensity fall-off. The next few most important nodes provide supporting information to effectively distinguish objects' effective radii. This sequential split becomes especially important to distinguish sequences in the smooth transition region where a few sequences from different classes show similar slopes. This sequential splitting as a qualitative measure of the slopes is physically meaningful: in fact, as commented already in Section \ref{sec:extraction_profiles}, we expect LTG sequences to fall off slower than ETG, due to their lower Sérsic index. As mentioned in Section \ref{sec:sampleselection} the PSF can change the rate of fall off from one isophote to the next, leading to a global rescaling of first steps of each sequence. This effect can be particularly significant for galaxies with size smaller than the PSF, but we did not include those in our sample.

\begin{figure}
	\includegraphics[width=\columnwidth]{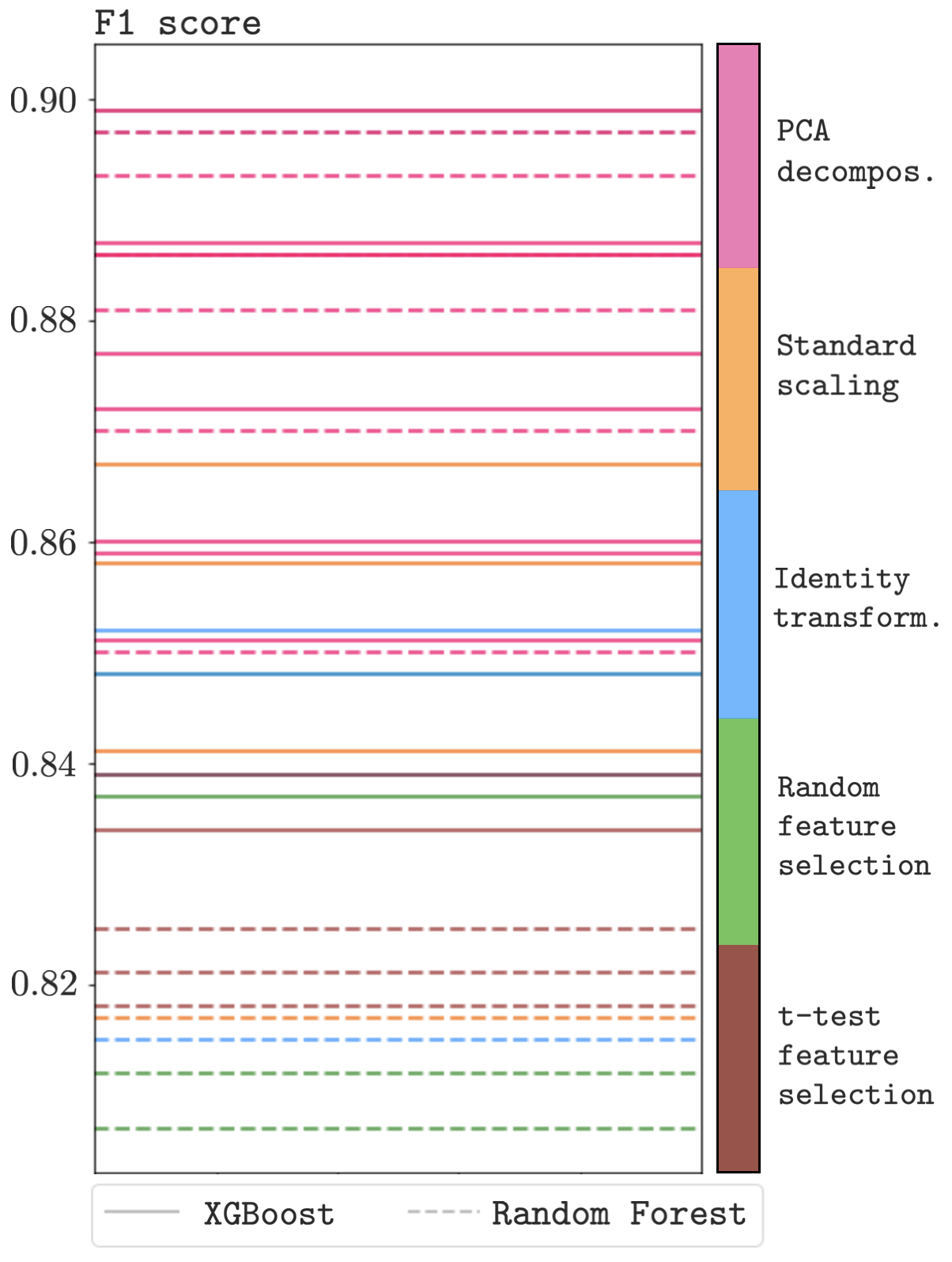}
	\caption{Summary of the results provided by the Modulos \texttt{AutoML} platform. The solutions are Decision Tree models, either XGBoost (solid lines) or Random Forest (dashed lines), which are able to sequentially capture and combine information from the sequences in the data set. The lines are color-coded by the feature engineering method. PCA decomposition is associated to the solutions maximizing the F1 score (in the y-axis), given their ability in rotating the feature space so to emphasize the most important nodes in the sequences.}
	\label{fig:results_summary}
\end{figure}

The scores of all trained solutions provided by the \texttt{AutoML} platform are summarized in Figure \ref{fig:results_summary}. Looking at these offers insights into which combinations of machine learning models and feature engineering methods are optimal for our task. The models shown are both decision trees-based and are either XGBoost (solid lines) or Random Forest models (dashed lines) and are color-coded by the feature engineering method. We observe that the PCA decomposition performs the best and is more important to the success of the overall model than the choice of XGBoost versus random forest. PCA rotates the feature space in order to successfully emphasize the slopes of the profiles. On the lower end we find the Random and the t-test feature selection methods: since they only select a subset of nodes (either random selection or by applying the Student's t-test), they seem to be less likely to pick up the most important information to distinguish the profiles and their slopes.

\begin{figure}
	\includegraphics[width=\columnwidth]{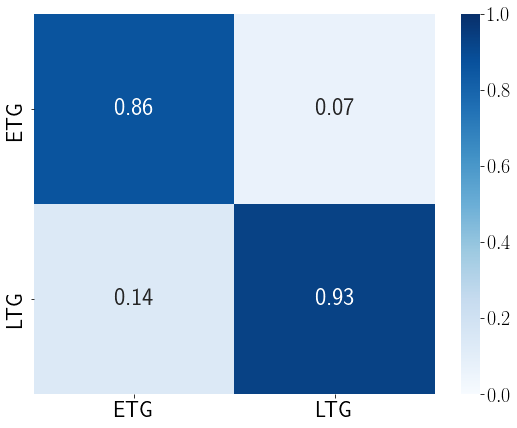}
	\caption{Confusion matrix representing the accuracy achieved in classifying galaxy profiles. The x-axis shows the true values, while the y-axis are the predicted categories. The main diagonal shows the correct classifications. The model seems quite robust in classifying the early-type galaxies of the sample.}
	\label{fig:confmatrix}
\end{figure}

\begin{figure*}
	\includegraphics[width=\textwidth]{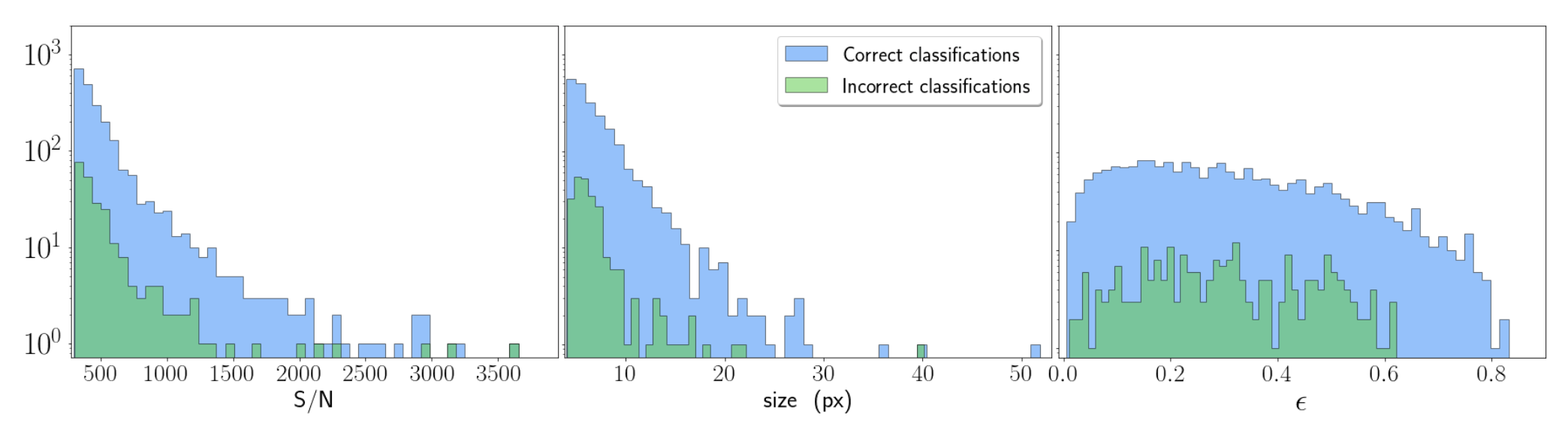}
	\caption{Properties of the classified sample in terms of signal-to-noise ratio (left panel), size (central panel) and ellipticity (right panel), distinguishing between objects with and without successful classification. This diagnostic plot alone cannot trace all the mis-classifications. A clearer test is shown in Figure \ref{fig:poorisofits}.}
	\label{fig:histoproperties}
\end{figure*}

Finally, we use the aforementioned best model to make predictions on our test sample. We quantify the distance between the predictions and the true values by computing the confusion matrix (Figure \ref{fig:confmatrix}), normalized over the number of predictions, for which we used the \texttt{Python sklearn} library. The main diagonal shows the amount of objects correctly classified, while the off-diagonal elements quantify incorrect classifications. The majority of mis-classified galaxies have low S/N ratios and tend to have small sizes and ellipticity, as shown in Figure \ref{fig:histoproperties}. 

\subsection{Model failures and future perspective}
\label{sec:altern}
\begin{figure*}
	\includegraphics[width=\textwidth]{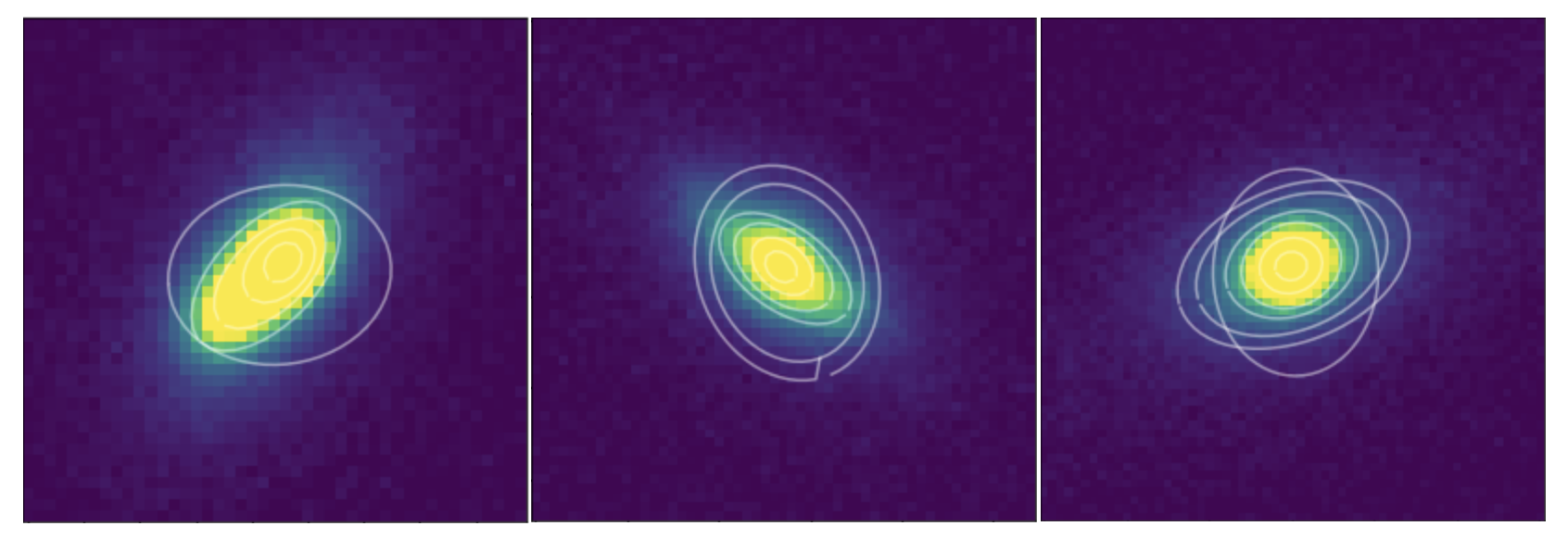}
	\caption{Examples of isophotal fitting for mis-classified galaxies. If compared to Fig.~\ref{fig:example_figure}, here we notice that the fitting fails at modelling galaxy wings and introduces rotations in the isophotal ellipses.}
	\label{fig:poorisofits}
\end{figure*}

Although there is by no means a simple cut we can perform to identify wrongly classified cases, inspecting examples of the isophotal fittings of both successful and unsuccessful classifications, we notice that objects with bad isophotal fitting tend to be mis-classified more often. This is compatible with the outcome shown in the confusion matrix, where it yields more incorrect classifications for ETG: a poor isophotal fit introduces perturbations into a 1D-sequence which would show a regular pattern typical of such galaxies. A few examples of galaxies with poor isophotal measurements are shown in Figure~\ref{fig:poorisofits}. Due to the small apparent size or low image resolution, the fitting does not model the light distribution well, resulting in an incorrect fit of the wings. As can be seen in the middle and right-hand panels, this manifests as a sudden change in the angular orientation of some isophotes with respect to the central regions of the galaxy. This issue can be corrected by applying sigma clipping to the recovered set of isophotes, identifying those that have parameters that are discrepant with the majority of fitted isophotes. However, the appropriate level of clipping varies from object to object and, at present, is not straight forward to determine in an automated way. As our aim is to describe a fully-automated method that can be run efficiently on large survey data, we thus quote our results without this fix. We will return to the issue of mis-aligned isophotes in future work, where we develop a routine to perform flexible isophotal fitting automatically, combining the structural information on the isophotes (e.g. position angle, ellipticity) with new feature engineering solutions, and apply our method to more contextual outputs, such as the presence of clumps or spiral arms. 

%Figure \ref{fig:poorisofits} shows the distribution of signal-to-noise ratio of the test sample. The light-blue histogram refers to the whole test sample, while the red one shows the objects with incorrect classification. Among those, the $45 \%$ have $S/N < 400$ and the $70 \%$ have $S/N < 500$. Further investigations made by looking at the corresponding images reveal several cases of poor signal-to-noise images.

\section{Conclusions}
\label{sec:discussion}
In this work, we describe a novel approach to galaxy morphological classification. It consists of first analysing the main features of the two-dimensional light distribution in a galaxy image with isophotal fitting. This then allows to unravel it to a one-dimensional sequence. The advantage of such an approach is the low complexity of one-dimensional data, which makes both data storage and processing easier and faster compared to classification methods directly analysing images (e.g. parametric fitting). The selection, calibration, and training of classification models is then performed using the Modulos \texttt{AutoML} platform, which allows users to intuitively build and run their workflows and automatizes the search and training of ML solutions. Using this platform also leads to a significant reduction of time spent on building machine learning algorithms. This allowed us to quickly test hypotheses and focus on the scientific analysis. We found ensembles of decision trees (XGBoost and Random Forest models) with a PCA decomposition as a feature engineering method, which transform the feature space to make the profiles more discrepant, to perform well. The resulting best performing model (XGBoost) is physically meaningful as it picks up on the differing slopes of the light profiles of galaxies: LTG profiles are expected to fall off slower due to their lower Sérsic index. 
We make the best ML solution we have found freely available at \href{https://github.com/Federica24/Cosmo}{https://github.com/Federica24/Cosmo}. It can be used to predict the galaxy type of other galaxies in the DES DR2 data set. We obtain an overall $F_1$ score of $90\%$ and $89\%$ on training and test data, respectively, which proves that the dimensionality reduction of the data, even though it implies information loss, still contains enough information to successfully classify galaxies. Our accuracy is comparable to the results found by \citep{vegaferrero2020pushing} for image-based classification using DES images. In the future, we will expand upon our promising results by developing a more robust isophotal measurement approach to focus on performance at low S/N, and target higher context features, such as bars, spiral arms and clumps.

\section*{Acknowledgements}

This research made use of Photutils, an Astropy package for
detection and photometry of astronomical sources \citep{photutils}. The authors thanks Modulos for the usage of their platform to perform image training and classification. 

This project used public archival data from the Dark Energy Survey (DES). Funding for the DES Projects has been provided by the U.S. Department of Energy, the U.S. National Science Foundation, the Ministry of Science and Education of Spain, the Science and Technology FacilitiesCouncil of the United Kingdom, the Higher Education Funding Council for England, the National Center for Supercomputing Applications at the University of Illinois at Urbana-Champaign, the Kavli Institute of Cosmological Physics at the University of Chicago, the Center for Cosmology and Astro-Particle Physics at the Ohio State University, the Mitchell Institute for Fundamental Physics and Astronomy at Texas A\&M University, Financiadora de Estudos e Projetos, Funda{\c c}{\~a}o Carlos Chagas Filho de Amparo {\`a} Pesquisa do Estado do Rio de Janeiro, Conselho Nacional de Desenvolvimento Cient{\'i}fico e Tecnol{\'o}gico and the Minist{\'e}rio da Ci{\^e}ncia, Tecnologia e Inova{\c c}{\~a}o, the Deutsche Forschungsgemeinschaft, and the Collaborating Institutions in the Dark Energy Survey.
The Collaborating Institutions are Argonne National Laboratory, the University of California at Santa Cruz, the University of Cambridge, Centro de Investigaciones Energ{\'e}ticas, Medioambientales y Tecnol{\'o}gicas-Madrid, the University of Chicago, University College London, the DES-Brazil Consortium, the University of Edinburgh, the Eidgen{\"o}ssische Technische Hochschule (ETH) Z{\"u}rich,  Fermi National Accelerator Laboratory, the University of Illinois at Urbana-Champaign, the Institut de Ci{\`e}ncies de l'Espai (IEEC/CSIC), the Institut de F{\'i}sica d'Altes Energies, Lawrence Berkeley National Laboratory, the Ludwig-Maximilians Universit{\"a}t M{\"u}nchen and the associated Excellence Cluster Universe, the University of Michigan, the National Optical Astronomy Observatory, the University of Nottingham, The Ohio State University, the OzDES Membership Consortium, the University of Pennsylvania, the University of Portsmouth, SLAC National Accelerator Laboratory, Stanford University, the University of Sussex, and Texas A\&M University.
Based in part on observations at Cerro Tololo Inter-American Observatory, National Optical Astronomy Observatory, which is operated by the Association of Universities for Research in Astronomy (AURA) under a cooperative agreement with the National Science Foundation.

%%%%%%%%%%%%%%%%%%%%%%%%%%%%%%%%%%%%%%%%%%%%%%%%%%
\section*{Data Availability}
The best automated classification model presented in this paper and discussed in Section \ref{sec:results} is publicly available at \href{https://github.com/Federica24/Cosmo}{https://github.com/Federica24/Cosmo} and can be used to classify any DES data from the public release DR2 processed with isophotal fitting, as described in this work.

%%%%%%%%%%%%%%%%%%%% REFERENCES %%%%%%%%%%%%%%%%%%

% The best way to enter references is to use BibTeX:

\bibliographystyle{mnras}
\bibliography{example} % if your bibtex file is called example.bib

\begin{thebibliography}{}
\makeatletter
\relax
\def\mn@urlcharsother{\let\do\@makeother \do\$\do\&\do\#\do\^\do\_\do\%\do\~}
\def\mn@doi{\begingroup\mn@urlcharsother \@ifnextchar [ {\mn@doi@}
  {\mn@doi@[]}}
\def\mn@doi@[#1]#2{\def\@tempa{#1}\ifx\@tempa\@empty \href
  {http://dx.doi.org/#2} {doi:#2}\else \href {http://dx.doi.org/#2} {#1}\fi
  \endgroup}
\def\mn@eprint#1#2{\mn@eprint@#1:#2::\@nil}
\def\mn@eprint@arXiv#1{\href {http://arxiv.org/abs/#1} {{\tt arXiv:#1}}}
\def\mn@eprint@dblp#1{\href {http://dblp.uni-trier.de/rec/bibtex/#1.xml}
  {dblp:#1}}
\def\mn@eprint@#1:#2:#3:#4\@nil{\def\@tempa {#1}\def\@tempb {#2}\def\@tempc
  {#3}\ifx \@tempc \@empty \let \@tempc \@tempb \let \@tempb \@tempa \fi \ifx
  \@tempb \@empty \def\@tempb {arXiv}\fi \@ifundefined
  {mn@eprint@\@tempb}{\@tempb:\@tempc}{\expandafter \expandafter \csname
  mn@eprint@\@tempb\endcsname \expandafter{\@tempc}}}

\bibitem[\protect\citeauthoryear{{Abbott} et~al.,}{{Abbott}
  et~al.}{2016}]{Abbott}
{Abbott} T.,  et~al., 2016, \mn@doi [\mnras] {10.1093/mnras/stw641}, \href
  {http://adsabs.harvard.edu/abs/2016MNRAS.460.1270D} {460, 1270}

\bibitem[\protect\citeauthoryear{Abbott et~al.,}{Abbott
  et~al.}{2021}]{abbott2021dark}
Abbott T. M.~C.,  et~al., 2021, The Dark Energy Survey Data Release 2
  (\mn@eprint {arXiv} {2101.05765})

\bibitem[\protect\citeauthoryear{Abraham, Aniyan, Kembhavi, Philip  \&
  Vaghmare}{Abraham et~al.}{2018}]{10.1093/mnras/sty627}
Abraham S.,  Aniyan A.~K.,  Kembhavi A.~K.,  Philip N.~S.,   Vaghmare K.,
  2018, \mn@doi [Monthly Notices of the Royal Astronomical Society]
  {10.1093/mnras/sty627}, 477, 894

\bibitem[\protect\citeauthoryear{Amiaux et~al.,}{Amiaux
  et~al.}{2012}]{Amiaux_2012}
Amiaux J.,  et~al., 2012, \mn@doi [Space Telescopes and Instrumentation 2012:
  Optical, Infrared, and Millimeter Wave] {10.1117/12.926513}

\bibitem[\protect\citeauthoryear{{Baldry}, {Balogh}, {Bower}, {Glazebrook}  \&
  {Nichol}}{{Baldry} et~al.}{2004}]{2004AIPC..743..106B}
{Baldry} I.~K.,  {Balogh} M.~L.,  {Bower} R.,  {Glazebrook} K.,   {Nichol}
  R.~C.,  2004, in {Allen} R.~E.,  {Nanopoulos} D.~V.,   {Pope} C.~N.,  eds,
  American Institute of Physics Conference Series Vol. 743, The New Cosmology:
  Conference on Strings and Cosmology. pp 106--119 (\mn@eprint {arXiv}
  {astro-ph/0410603}), \mn@doi{10.1063/1.1848322}

\bibitem[\protect\citeauthoryear{{Banerji} et~al.,}{{Banerji}
  et~al.}{2010}]{Manda2010}
{Banerji} M.,  et~al., 2010, \mn@doi [\mnras]
  {10.1111/j.1365-2966.2010.16713.x}, \href
  {http://adsabs.harvard.edu/abs/2010MNRAS.406..342B} {406, 342}

\bibitem[\protect\citeauthoryear{{Bertin}}{{Bertin}}{2011}]{Bertin2011}
{Bertin} E.,  2011, in {Evans} I.~N.,  {Accomazzi} A.,  {Mink} D.~J.,   {Rots}
  A.~H.,  eds,  Astronomical Society of the Pacific Conference Series Vol. 442,
  Astronomical Data Analysis Software and Systems XX. p.~435

\bibitem[\protect\citeauthoryear{Bradley et~al.,}{Bradley
  et~al.}{2020}]{photutils}
Bradley L.,  et~al., 2020, astropy/photutils: 1.0.0,
  \mn@doi{10.5281/zenodo.4044744}, \url
  {https://doi.org/10.5281/zenodo.4044744}

\bibitem[\protect\citeauthoryear{Cano-Díaz, Ávila Reese, Sánchez,
  Hernández-Toledo, Rodríguez-Puebla, Boquien  \& Ibarra-Medel}{Cano-Díaz
  et~al.}{2019}]{10.1093/mnras/stz1894}
Cano-Díaz M.,  Ávila Reese V.,  Sánchez S.~F.,  Hernández-Toledo H.~M.,
  Rodríguez-Puebla A.,  Boquien M.,   Ibarra-Medel H.,  2019, \mn@doi [Monthly
  Notices of the Royal Astronomical Society] {10.1093/mnras/stz1894}, 488, 3929

\bibitem[\protect\citeauthoryear{{Cheng}, {Huertas-Company}, {Conselice},
  {Arag{\'o}n-Salamanca}, {Robertson}  \& {Ramachandra}}{{Cheng}
  et~al.}{2020}]{2020arXiv200911932C}
{Cheng} T.-Y.,  {Huertas-Company} M.,  {Conselice} C.~J.,
  {Arag{\'o}n-Salamanca} A.,  {Robertson} B.~E.,   {Ramachandra} N.,  2020,
  arXiv e-prints, \href {https://ui.adsabs.harvard.edu/abs/2020arXiv200911932C}
  {p. arXiv:2009.11932}

\bibitem[\protect\citeauthoryear{{Combes} \& {Sanders}}{{Combes} \&
  {Sanders}}{1981}]{1981A&A....96..164C}
{Combes} F.,  {Sanders} R.~H.,  1981, \aap, \href
  {https://ui.adsabs.harvard.edu/abs/1981A&A....96..164C} {96, 164}

\bibitem[\protect\citeauthoryear{{Conselice}, {Bershady}  \&
  {Jangren}}{{Conselice} et~al.}{2000}]{Conselice2000}
{Conselice} C.~J.,  {Bershady} M.~A.,   {Jangren} A.,  2000, \mn@doi [\apj]
  {10.1086/308300}, \href {http://adsabs.harvard.edu/abs/2000ApJ...529..886C}
  {529, 886}

\bibitem[\protect\citeauthoryear{{Dark Energy Survey Collaboration}
  et~al.,}{{Dark Energy Survey Collaboration}
  et~al.}{2016}]{2016MNRAS.460.1270D}
{Dark Energy Survey Collaboration} et~al., 2016, \mn@doi [\mnras]
  {10.1093/mnras/stw641}, \href
  {https://ui.adsabs.harvard.edu/abs/2016MNRAS.460.1270D} {460, 1270}

\bibitem[\protect\citeauthoryear{{Desai} et~al.,}{{Desai} et~al.}{2012}]{Desai}
{Desai} S.,  et~al., 2012, \mn@doi [\apj] {10.1088/0004-637X/757/1/83}, \href
  {http://adsabs.harvard.edu/abs/2012ApJ...757...83D} {757, 83}

\bibitem[\protect\citeauthoryear{{Dieleman}, {Willett}  \& {Dambre}}{{Dieleman}
  et~al.}{2015}]{Dieleman}
{Dieleman} S.,  {Willett} K.~W.,   {Dambre} J.,  2015, \mn@doi [\mnras]
  {10.1093/mnras/stv632}, \href
  {http://adsabs.harvard.edu/abs/2015MNRAS.450.1441D} {450, 1441}

\bibitem[\protect\citeauthoryear{{Drlica-Wagner} et~al.,}{{Drlica-Wagner}
  et~al.}{2017}]{Drlica-Wagner}
{Drlica-Wagner} A.,  et~al., 2017, preprint, \href
  {http://adsabs.harvard.edu/abs/2017arXiv170801531D} {} (\mn@eprint {arXiv}
  {1708.01531})

\bibitem[\protect\citeauthoryear{Drlica-Wagner et~al.,}{Drlica-Wagner
  et~al.}{2018}]{Drlica_Wagner_2018}
Drlica-Wagner A.,  et~al., 2018, \mn@doi [The Astrophysical Journal Supplement
  Series] {10.3847/1538-4365/aab4f5}, 235, 33

\bibitem[\protect\citeauthoryear{{Elmegreen}, {Elmegreen}, {Chromey},
  {Hasselbacher}  \& {Bissell}}{{Elmegreen} et~al.}{1996}]{1996AJ....111.2233E}
{Elmegreen} B.~G.,  {Elmegreen} D.~M.,  {Chromey} F.~R.,  {Hasselbacher} D.~A.,
    {Bissell} B.~A.,  1996, \mn@doi [\aj] {10.1086/117957}, \href
  {https://ui.adsabs.harvard.edu/abs/1996AJ....111.2233E} {111, 2233}

\bibitem[\protect\citeauthoryear{{Flaugher}}{{Flaugher}}{2005}]{Flaugher}
{Flaugher} B.,  2005, \mn@doi [International Journal of Modern Physics A]
  {10.1142/S0217751X05025917}, \href
  {http://adsabs.harvard.edu/abs/2005IJMPA..20.3121F} {20, 3121}

\bibitem[\protect\citeauthoryear{{Flaugher} et~al.,}{{Flaugher}
  et~al.}{2015a}]{2015AJ....150..150F}
{Flaugher} B.,  et~al., 2015a, \mn@doi [\aj] {10.1088/0004-6256/150/5/150},
  \href {https://ui.adsabs.harvard.edu/abs/2015AJ....150..150F} {150, 150}

\bibitem[\protect\citeauthoryear{Flaugher et~al.,}{Flaugher
  et~al.}{2015b}]{Flaugher_2015}
Flaugher B.,  et~al., 2015b, \mn@doi [The Astronomical Journal]
  {10.1088/0004-6256/150/5/150}, 150, 150

\bibitem[\protect\citeauthoryear{Friedman}{Friedman}{2001}]{10.1214/aos/1013203451}
Friedman J.~H.,  2001, \mn@doi [The Annals of Statistics]
  {10.1214/aos/1013203451}, 29, 1189

\bibitem[\protect\citeauthoryear{{Ghosh}, {Urry}, {Wang}, {Schawinski}, {Turp}
  \& {Powell}}{{Ghosh} et~al.}{2020}]{2020ApJ...895..112G}
{Ghosh} A.,  {Urry} C.~M.,  {Wang} Z.,  {Schawinski} K.,  {Turp} D.,   {Powell}
  M.~C.,  2020, \mn@doi [\apj] {10.3847/1538-4357/ab8a47}, \href
  {https://ui.adsabs.harvard.edu/abs/2020ApJ...895..112G} {895, 112}

\bibitem[\protect\citeauthoryear{Goncalves, Martin, Menéndez-Delmestre, Wyder
  \& Koekemoer}{Goncalves et~al.}{2012}]{article}
Goncalves T.,  Martin C.,  Menéndez-Delmestre K.,  Wyder T.,   Koekemoer A.,
  2012, \mn@doi [Proceedings of the International Astronomical Union]
  {10.1017/S1743921313004572}, 8, 163

\bibitem[\protect\citeauthoryear{Grogin et~al.,}{Grogin
  et~al.}{2011}]{Grogin_2011}
Grogin N.~A.,  et~al., 2011, \mn@doi [The Astrophysical Journal Supplement
  Series] {10.1088/0067-0049/197/2/35}, 197, 35

\bibitem[\protect\citeauthoryear{Huertas-Company et~al.,}{Huertas-Company
  et~al.}{2015}]{Huertas_Company_2015}
Huertas-Company M.,  et~al., 2015, \mn@doi [The Astrophysical Journal
  Supplement Series] {10.1088/0067-0049/221/1/8}, 221, 8

\bibitem[\protect\citeauthoryear{{Jedrzejewski}}{{Jedrzejewski}}{1987}]{isophotes}
{Jedrzejewski} R.~I.,  1987, \mn@doi [\mnras] {10.1093/mnras/226.4.747}, \href
  {https://ui.adsabs.harvard.edu/abs/1987MNRAS.226..747J} {226, 747}

\bibitem[\protect\citeauthoryear{Koekemoer et~al.,}{Koekemoer
  et~al.}{2011}]{Koekemoer_2011}
Koekemoer A.~M.,  et~al., 2011, \mn@doi [The Astrophysical Journal Supplement
  Series] {10.1088/0067-0049/197/2/36}, 197, 36

\bibitem[\protect\citeauthoryear{{Lingard} et~al.,}{{Lingard}
  et~al.}{2020}]{Lingard}
{Lingard} T.~K.,  et~al., 2020, \mn@doi [\apj] {10.3847/1538-4357/ab9d83},
  \href {https://ui.adsabs.harvard.edu/abs/2020ApJ...900..178L} {900, 178}

\bibitem[\protect\citeauthoryear{{Lintott} et~al.,}{{Lintott}
  et~al.}{2008}]{LintottKevin}
{Lintott} C.~J.,  et~al., 2008, \mn@doi [\mnras]
  {10.1111/j.1365-2966.2008.13689.x}, \href
  {http://adsabs.harvard.edu/abs/2008MNRAS.389.1179L} {389, 1179}

\bibitem[\protect\citeauthoryear{{Mohr} et~al.,}{{Mohr} et~al.}{2012}]{Mohr}
{Mohr} J.~J.,  et~al., 2012, in Software and Cyberinfrastructure for Astronomy
  II. p. 84510D (\mn@eprint {arXiv} {1207.3189}), \mn@doi{10.1117/12.926785}

\bibitem[\protect\citeauthoryear{{Morganson} et~al.,}{{Morganson}
  et~al.}{2018a}]{Morganson2018}
{Morganson} E.,  et~al., 2018a, preprint, \href
  {http://adsabs.harvard.edu/abs/2018arXiv180103177M} {} (\mn@eprint {arXiv}
  {1801.03177})

\bibitem[\protect\citeauthoryear{{Morganson} et~al.,}{{Morganson}
  et~al.}{2018b}]{2018PASP..130g4501M}
{Morganson} E.,  et~al., 2018b, \mn@doi [\pasp] {10.1088/1538-3873/aab4ef},
  \href {https://ui.adsabs.harvard.edu/abs/2018PASP..130g4501M} {130, 074501}

\bibitem[\protect\citeauthoryear{Nair \& Abraham}{Nair \&
  Abraham}{2010}]{Nair_2010}
Nair P.~B.,  Abraham R.~G.,  2010, \mn@doi [The Astrophysical Journal
  Supplement Series] {10.1088/0067-0049/186/2/427}, 186, 427

\bibitem[\protect\citeauthoryear{{Noeske} et~al.,}{{Noeske}
  et~al.}{2007}]{2007ApJ...660L..47N}
{Noeske} K.~G.,  et~al., 2007, \mn@doi [\apjl] {10.1086/517927}, \href
  {https://ui.adsabs.harvard.edu/abs/2007ApJ...660L..47N} {660, L47}

\bibitem[\protect\citeauthoryear{{Palmese} et~al.,}{{Palmese}
  et~al.}{2017}]{Palmese}
{Palmese} A.,  et~al., 2017, \mn@doi [\apjl] {10.3847/2041-8213/aa9660}, \href
  {https://ui.adsabs.harvard.edu/abs/2017ApJ...849L..34P} {849, L34}

\bibitem[\protect\citeauthoryear{{Peng}, {Ho}, {Impey}  \& {Rix}}{{Peng}
  et~al.}{2010}]{galfit}
{Peng} C.~Y.,  {Ho} L.~C.,  {Impey} C.~D.,   {Rix} H.-W.,  2010, \mn@doi [\aj]
  {10.1088/0004-6256/139/6/2097}, \href
  {https://ui.adsabs.harvard.edu/abs/2010AJ....139.2097P} {139, 2097}

\bibitem[\protect\citeauthoryear{{Peterken}, {Merrifield},
  {Arag{\'o}n-Salamanca}, {Avila-Reese}, {Boardman}, {Drory}  \&
  {Lane}}{{Peterken} et~al.}{2021}]{2021MNRAS.500L..42P}
{Peterken} T.,  {Merrifield} M.,  {Arag{\'o}n-Salamanca} A.,  {Avila-Reese} V.,
   {Boardman} N.~F.,  {Drory} N.,   {Lane} R.~R.,  2021, \mn@doi [\mnras]
  {10.1093/mnrasl/slaa179}, \href
  {https://ui.adsabs.harvard.edu/abs/2021MNRAS.500L..42P} {500, L42}

\bibitem[\protect\citeauthoryear{Salim}{Salim}{2014}]{Salim_2014}
Salim S.,  2014, \mn@doi [Serbian Astronomical Journal] {10.2298/saj1489001s},
  p. 1–14

\bibitem[\protect\citeauthoryear{Schawinski et~al.,}{Schawinski
  et~al.}{2014}]{Schawinski_2014}
Schawinski K.,  et~al., 2014, \mn@doi [Monthly Notices of the Royal
  Astronomical Society] {10.1093/mnras/stu327}, 440, 889–907

\bibitem[\protect\citeauthoryear{{Schiminovich} et~al.,}{{Schiminovich}
  et~al.}{2007}]{2007ApJS..173..315S}
{Schiminovich} D.,  et~al., 2007, \mn@doi [\apjs] {10.1086/524659}, \href
  {https://ui.adsabs.harvard.edu/abs/2007ApJS..173..315S} {173, 315}

\bibitem[\protect\citeauthoryear{{S{\'e}rsic}}{{S{\'e}rsic}}{1963}]{Sersic}
{S{\'e}rsic} J.~L.,  1963, Boletin de la Asociacion Argentina de Astronomia La
  Plata Argentina, \href {http://adsabs.harvard.edu/abs/1963BAAA....6...41S}
  {6, 41}

\bibitem[\protect\citeauthoryear{Sevilla et~al.}{Sevilla
  et~al.}{2011}]{Sevilla:2011ps}
Sevilla I.,  et~al., 2011, in {Particles and fields. Proceedings, Meeting of
  the Division of the American Physical Society, DPF 2011, Providence, USA,
  August 9-13, 2011}.  (\mn@eprint {arXiv} {1109.6741}), \url
  {http://lss.fnal.gov/cgi-bin/find_paper.pl?conf-11-525}

\bibitem[\protect\citeauthoryear{{Simard}, {Mendel}, {Patton}, {Ellison}  \&
  {McConnachie}}{{Simard} et~al.}{2011}]{Simard2011}
{Simard} L.,  {Mendel} J.~T.,  {Patton} D.~R.,  {Ellison} S.~L.,
  {McConnachie} A.~W.,  2011, \mn@doi [\apjs] {10.1088/0067-0049/196/1/11},
  \href {http://adsabs.harvard.edu/abs/2011ApJS..196...11S} {196, 11}

\bibitem[\protect\citeauthoryear{{Simmons} et~al.,}{{Simmons}
  et~al.}{2017}]{Simmons}
{Simmons} B.~D.,  et~al., 2017, \mn@doi [\mnras] {10.1093/mnras/stw2587}, \href
  {http://adsabs.harvard.edu/abs/2017MNRAS.464.4420S} {464, 4420}

\bibitem[\protect\citeauthoryear{{Srinivas}, {Krause}, {Kakade}  \&
  {Seeger}}{{Srinivas} et~al.}{2009}]{srinivas:2009aa}
{Srinivas} N.,  {Krause} A.,  {Kakade} S.~M.,   {Seeger} M.,  2009, arXiv
  e-prints, \href {https://ui.adsabs.harvard.edu/abs/2009arXiv0912.3995S} {p.
  arXiv:0912.3995}

\bibitem[\protect\citeauthoryear{{Tarsitano} et~al.,}{{Tarsitano}
  et~al.}{2018}]{catdes}
{Tarsitano} F.,  et~al., 2018, \mn@doi [\mnras] {10.1093/mnras/sty1970}, \href
  {https://ui.adsabs.harvard.edu/abs/2018MNRAS.481.2018T} {481, 2018}

\bibitem[\protect\citeauthoryear{{Tuccillo}, {Huertas-Company},
  {Decenci{\`e}re}, {Velasco-Forero}, {Dom{\'\i}nguez S{\'a}nchez}  \&
  {Dimauro}}{{Tuccillo} et~al.}{2018}]{2018MNRAS.475..894T}
{Tuccillo} D.,  {Huertas-Company} M.,  {Decenci{\`e}re} E.,  {Velasco-Forero}
  S.,  {Dom{\'\i}nguez S{\'a}nchez} H.,   {Dimauro} P.,  2018, \mn@doi [\mnras]
  {10.1093/mnras/stx3186}, \href
  {https://ui.adsabs.harvard.edu/abs/2018MNRAS.475..894T} {475, 894}

\bibitem[\protect\citeauthoryear{Vega-Ferrero et~al.,}{Vega-Ferrero
  et~al.}{2020}]{vegaferrero2020pushing}
Vega-Ferrero J.,  et~al., 2020, Pushing automated morphological classifications
  to their limits with the Dark Energy Survey (\mn@eprint {arXiv} {2012.07858})

\bibitem[\protect\citeauthoryear{Walmsley et~al.,}{Walmsley
  et~al.}{2019}]{10.1093/mnras/stz2816}
Walmsley M.,  et~al., 2019, \mn@doi [Monthly Notices of the Royal Astronomical
  Society] {10.1093/mnras/stz2816}, 491, 1554

\bibitem[\protect\citeauthoryear{{York} et~al.,}{{York}
  et~al.}{2000}]{2000AJ....120.1579Y}
{York} D.~G.,  et~al., 2000, \mn@doi [\aj] {10.1086/301513}, \href
  {https://ui.adsabs.harvard.edu/abs/2000AJ....120.1579Y} {120, 1579}

\bibitem[\protect\citeauthoryear{{de Jong}}{{de
  Jong}}{1996}]{1996A&A...313...45D}
{de Jong} R.~S.,  1996, \aap, \href
  {https://ui.adsabs.harvard.edu/abs/1996A&A...313...45D} {313, 45}

\makeatother
\end{thebibliography}

% Alternatively you could enter them by hand, like this:
% This method is tedious and prone to error if you have lots of references
%\begin{thebibliography}{99}
%\bibitem[\protect\citeauthoryear{Author}{2012}]{Author2012}
%Author A.~N., 2013, Journal of Improbable Astronomy, 1, 1
%\bibitem[\protect\citeauthoryear{Others}{2013}]{Others2013}
%Others S., 2012, Journal of Interesting Stuff, 17, 198
%\end{thebibliography}

%%%%%%%%%%%%%%%%%%%%%%%%%%%%%%%%%%%%%%%%%%%%%%%%%%

%%%%%%%%%%%%%%%%% APPENDICES %%%%%%%%%%%%%%%%%%%%%

%\appendix

%\section{Some extra material}

%If you want to present additional material which would interrupt the flow of the main paper,
%it can be placed in an Appendix which appears after the list of references.

%%%%%%%%%%%%%%%%%%%%%%%%%%%%%%%%%%%%%%%%%%%%%%%%%%

% Don't change these lines
\bsp	% typesetting comment
\label{lastpage}
\end{document}